\documentstyle[12pt,epsf,epsfig]{article}
\def\pgb{{pseudo-Goldstone bosons}}
\def\ppgb{{Pseudo-Goldstone bosons}}
\def\beq{\begin{equation}}
\def\eeq{\end{equation}}
\def\bea{\begin{eqnarray}}
\def\eea{\end{eqnarray}}
\def\ba{\begin{array}}
\def\ea{\end{array}}
\def\gappeq{\mathrel{\rlap {\raise.5ex\hbox{$>$}}
{\lower.5ex\hbox{$\sim$}}}}
\def\permil{$\%\raise.20ex\hbox{$_0$}}
\def\lappeq{\mathrel{\rlap{\raise.5ex\hbox{$<$}}
{\lower.5ex\hbox{$\sim$}}}}
\begin{document}
\topmargin -1.0cm
\oddsidemargin -0.8cm
\evensidemargin -0.8cm
\pagestyle{empty}
\begin{flushright}
CERN-TH/96-47
\end{flushright}
\vspace*{5mm}
\begin{center}
{\Large \bf Macroscopic Forces from Supersymmetry\footnote{Based on
talk given
at the Santa Barbara Workshop on SusyGUTs, December 6-9, 1995.}}\\
\vspace{1.5cm}
{\large S. Dimopoulos$^{\#,\ast}$ and
G.F. Giudice$^\#$\footnote{On leave of absence from INFN, Sezione di
Padova,
Padua, Italy.}}\\
\vspace{0.3cm}
$^\#$Theoretical Physics Division, CERN\\
CH-1211 Geneva 23, Switzerland\\
\vspace{0.3cm}
$^\ast$Physics Department, Stanford University\\ Stanford CA 94305,
USA\\
\vspace*{2cm}
Abstract
\end{center}
We argue that theories in which supersymmetry breaking originates at
low energies often contain scalar particles that mediate coherent
gravitational strength forces at distances less than a cm. We
estimate
the strength and range of these forces in several cases.  Present
limits on such forces are inadequate. However new techniques,
such as those based on small cryogenic mechanical oscillators, may
improve the present limits by ten orders of magnitude or discover new
forces as weak as 1 \% of gravity at distances down to 40
microns.

\vfill
\begin{flushleft}
CERN-TH/96-47\\
February 1996
\end{flushleft}
\eject
\pagestyle{empty}
\setcounter{page}{1}
\setcounter{footnote}{0}
\pagestyle{plain}

\section{Light Scalars and Low-Energy Supersymmetry Breaking}
 
A trademark of superstring theories is the occurrence of
gravitationally coupled massles scalars called moduli. To avoid
conflict with Newtonean gravity moduli must obtain mass. One
possibility is that stringy non-perturbative phenomena create a
potential which gives them a  mass $\sim M_{\rm PL}$. A second
possibility is that Planckean physics leaves the moduli massless
and they obtain mass only as a result of
supersymmetry breaking. In
this case, since they are gravitationally
coupled, they are expected to get a mass proportional to  $\sim
F/M_{\rm PL}$,
where $F$ is the
scale where supersymmetry breaking originates. In theories with
gravity-mediated
supersymmetry breaking $F$ is about $(10^{11}~{\rm GeV})^2$ and
the moduli masses are at the weak scale.
Since they are only gravitationally coupled, they are not directly
relevant for experiment. Recently the supersymmetry flavour
problem \cite{ds} has
renewed interest in theories with low-energy
supersymmetry breaking\footnote{The possibility that supersymmetry is
broken at low energies with gauge interactions serving as messengers
was considered by a number of authors in the early 1980's. These
models are reviewed in \cite{nil}. More recently, models with
dynamical
breaking of supersymmetry at low energies have been proposed in
ref.~\cite{gaug}.},
where $F$ can be as small as $(10~{\rm TeV})^2$.
Moduli in these theories are so light that they can have macroscopic
Compton
wavelenghts and mediate macroscopic forces of gravitational strength.
In
this paper we estimate the range and magnitude of the force mediated
by
moduli in theories where supersymmetry breaking originates at low
energies. We separate the moduli in  two  categories: the Yukawa
moduli, which determine the Yukawa couplings, and the gauge moduli
which fix the gauge couplings. We find that both types of moduli can
mediate potentially measurable forces especially in the millimeter
range.

In addition we study, in section 3, the forces mediated by
(pseudo-)Goldstone bosons of the broken flavour group. We find that
these forces are potentially observable only if the flavour group is
broken at energies smaller than $M_{GUT}$.
In the last section we comment on the present limits and a possible
future  search for sub-cm range forces; these results are summarized
in fig.~2 together with some of our predictions. Our results for
the range and magnitude of the moduli and Goldstone mediated forces
are summarized in the tables.

\section{Moduli}
 
All the parameters of the supersymmetric standard model ---Yukawas,
gauge couplings, soft terms, the $\mu$ term--- may depend on moduli
which are undetermined by Planckean physics.
All such moduli can mediate macroscopic forces. In this paper we
shall focus on two classes of moduli which are of special interest
because they couple directly to ordinary matter and therefore mediate
the strongest forces. These are the Yukawa moduli and the gauge
moduli.
Consider  a Yukawa modulus $\phi$, coupled as follows to up-type
quarks and the Higgs boson $H$:
 
 
\beq
{\cal L}=\lambda (\phi )q_L {\bar u}_R H + {\rm h.c.}
\label{int}
\eeq
For definitness we assume that the field $\phi$ appears only in a
single,
but arbitrary, Yukawa
coupling and we drop the flavour indices. Interactions analogous to
eq.~(\ref{int}) can also occur for down-type quarks and charged
leptons.
As discussed in sect.~1, $\phi$ is a field with
Planck-suppressed
couplings which is postulated to have no potential until
supersymmetry
breaking turns on.

Supersymmetry breaking is expected to give $\phi$ a mass
proportional to the gravitino mass, $m_\phi\sim F/M_{\rm PL}$, with a
coefficient that depends on the couplings of $\phi$ to other fields.
We do not know the complete form of the effective potential $V(\phi
)$,
but at least we can compute the contribution coming from the operator
in eq.~(\ref{int}). Including explicitly
the dominant contribution from fig.~1a ,
$V(\phi )$ is given by
\beq
V(\phi )= \frac{k}{(16 \pi^2)^2}\lambda^\dagger (\phi )\lambda (\phi
)
m_s^2\Lambda^2+V_0(\phi )~.
\label{pot}
\eeq
Here $V_0(\phi )$ is
the unknown part of the potential and $k$ is a coefficient,
expected to be of order one, which parametrizes the loop integration.
$\Lambda$ is the scale above which the soft terms shut off and
supersymmetry
is recovered. In ordinary gravity-mediated supersymmetry breaking
$\Lambda\sim M_{\rm PL}$. In theories with low-energy gauge-mediated
supersymmetry breaking $\Lambda$ is of the order of the mass of the
messengers which communicate supersymmetry breaking to the ordinary
sector.
Contributions from figs.~1b-1c are typically less important, as they
have at
most a logatithmic dependence on $\Lambda$.
Finally, $m_s^2$ is a measure of supersymmetry-breaking
in the ordinary particle sector; it is the larger between the squark
mass squared and the higgs-higgsino mass squared splitting. For
definitness we assume it is equal to the squark mass squared which,
in
models where supersymmetry breaking is communicated by gauge
interactions,
can be related to the messenger mass $\Lambda$ \cite{gaug,alv}:
\beq
m_{\tilde q}^2\simeq \frac{8N}{3}\left(\frac{\alpha_s}{4\pi}\right)^2
\frac{F^2}{\Lambda^2}~.
\label{mass}
\eeq
Here $N$ is the number of messenger multiplets and $F$ is the scale
where
supersymmetry breaking originates or, more precisely, a measure of
the
messengers' mass splittings.

The requirement that all gauge couplings
remain perturbative below the GUT scale implies $N<4$. A lower bound
on the scale $\Lambda$ can be obtained from the experimental limit
on the right-handed selectron mass,
\beq
m_{{\tilde e}_R}^2\simeq
\frac{10N}{3}\left(\frac{\alpha}{4\pi\cos^2\theta_W}\right)^2
\frac{F^2}{\Lambda^2}~,
\eeq
and the consistency condition that messengers do not receive a
negative
mass squared, {\it i.e.} $F<\Lambda^2$. We obtain that $\Lambda$ can
be
as small as $30/\sqrt{N}$ TeV.

Replacing $m_s^2$ in eq.~(\ref{pot}) with eq.~(\ref{mass}),
we find
\beq
V(\phi )= \frac{8kN\alpha_s^2}{3(16 \pi^2)^3}\lambda^\dagger (\phi
)\lambda (\phi )
F^2+V_0(\phi )~.
\label{scass}
\eeq
To extract the mass of $\phi$, it is convenient to Taylor-expand
$\lambda
(\phi )$ around its minimum $\langle \phi \rangle \sim M$, where $M$
is
expected to be of the order of the string scale $5\times 10^{17}$
GeV,
\beq
\lambda (\phi )=\lambda^{(0)}+\lambda^{(1)}\frac{(\phi -\langle \phi
\rangle
)}{M} +\frac{1}{2}\lambda^{(2)}\frac{(\phi -\langle \phi \rangle
)^2}{M^2}+...~,
\eeq
\beq
\lambda^{(i)}\equiv \frac{d^i\lambda}{d\phi^i}(\phi =\langle \phi
\rangle )~.
\eeq
The $\phi$ mass is now extracted from eq.~(\ref{scass})
\beq
m_\phi^2=
\frac{16kN\alpha_s^2}{3(16
\pi^2)^3}(\lambda^{(1)2}+\lambda^{(0)}\lambda^{(2)})
\frac{F^2}{M^2}+\frac{d^2 V_0}{d\phi^2}(\phi =\langle \phi \rangle
)~,
\label{phi}
\eeq
and the minimization of the potential implies:
\beq
\lambda^{(1)}=
-\frac{3(16 \pi^2)^3}{16kN\alpha_s^2\lambda^{(0)}}
\frac{M_{\rm PL}}{F^2}~\frac{d V_0}{d\phi}(\phi =\langle \phi \rangle
)~.
\eeq
Assuming no accidental cancellation among the different terms in
eq.~(\ref{phi}), from the known part of the potential
we can at least establish a lower bound on the $\phi$-mass
or an upper bound on its Compton wavelength
\beq
\lambda_\phi \simeq 90~ \mu{\rm m}~\frac{(100~{\rm TeV})^2}{F}~
\left( \frac{M}{5\times 10^{17}~{\rm GeV}}\right)~
\left(
\frac{{\rm GeV}}{m_q}\right)\left( \frac{H_q}{1/\sqrt{2}}\right)
\left[ kN
\left(\frac{\lambda^{(1)2}}{\lambda^{(0)2}}
+\frac{\lambda^{(2)}}{\lambda^{(0)}}\right) \right]^{-1/2}~.
\eeq
Here $m_q$ is the mass of the corresponding quark and $H_q$ is equal
to
$\sin\beta$ for up-type quarks and $\cos\beta$ for down-type quarks,
where
$\tan\beta$ is the ratio of the two Higgs vacuum expectation values.
 
In order to compute the long-range force potential, we first need to
relate
the scalar $\phi$-quark coupling of eq.~(\ref{int}) to the scalar
$\phi$-nucleon coupling. The matrix element of the light-quark
current
can be obtained from the measurements of the pion-nucleon ``sigma
term".
However, to extract the separate matrix elements for up and down
quarks,
we need to rely on particular model calculations. Following
ref.~\cite{cina}, we take the
proton and neutron matrix elements to be
\beq
\langle p | m_d {\bar d} d |p\rangle = 0.034 ~m_N ~~~~~~
\langle n | m_d {\bar d} d |n\rangle = 0.041 ~m_N
\label{matd}
\eeq
\beq
\langle p | m_u {\bar u} u |p\rangle = 0.023 ~m_N ~~~~~~
\langle n | m_u {\bar u} u |n\rangle = 0.019 ~m_N
\label{matu}
\eeq
where $m_N$ is the nucleon mass.
It is interesting to notice that the fields $\phi$ corresponding to
up and down quarks have different couplings
to the proton and the neutron.
This leads to forces which depend on the atomic number of the
test material and therefore to small violations of the equivalence
principle
at macroscopic scales.
 
The strange-quark current has a larger matrix element. We use the
result
from ref.~\cite{leut}
\beq
\langle N | m_s {\bar s} s |N\rangle = 0.14 ~m_N ~,
\label{mats}
\eeq
although values as large as $0.4~m_N$ \cite{cina} and as small as
$0.08~m_N$
\cite{giap} are quoted in the literature.
 
The matrix element of the heavy-quark current can be computed by
relating it
to the gluon matrix element through the anomaly \cite{shif}. Finally
the
gluon matrix element is computed by expressing the nucleon mass in
terms
of the trace of the QCD energy-momentum tensor:
\beq
\langle N | m_Q {\bar Q} Q |N\rangle =
\frac{2}{27}\left(m_N-\sum_{q=u,d,s}
\langle N | m_q {\bar q} q |N\rangle \right) = 0.06~m_N ~~~~~~
\label{matq}
\eeq
Equation (\ref{matq}) holds for each heavy quark, $Q=c,b,t$.
 
The scalar coupling of the field $\phi$ to the nucleon $N$ is
\beq
{\cal L}={\cal G}_{\phi}\frac{m_N}{M_{\rm PL}} \phi {\bar \psi}_N
\psi_N~,
\label{couppp}
\eeq
where
\beq
{\cal G}_\phi = \frac{\lambda^{(1)}}{\lambda^{(0)}}~\frac{M_{\rm
PL}}{M}~
\frac{\langle N | m_q
{\bar q} q |N\rangle }{m_N}~.
\eeq
From eq.~(\ref{couppp}) we can now derive the potential
between two particles with masses $m_1$ and $m_2$ at a
distance $r$, which is generated
by a one-$\phi$ exchange:
\beq
V(r)=G_N m_1m_2{\cal G}_\phi^2\int
\frac{d^3k}{(2\pi)^3}\frac{e^{i\vec{k}
\cdot \vec{r}}}{\vec{k}^2+m_\phi^2}~.
\label{vecpot}
\eeq
When added to gravity, eq.~(\ref{vecpot}) describes an additional
attractive
force:
\beq
V(r) = - G_N \frac{m_1m_2}{r}\left( 1+\frac{{\cal G}_{\phi}^2}{4
\pi}e^{-r/\lambda_{\phi}}\right)~.
\eeq
 
The results for the Compton wavelengths and the strengths of the
moduli forces
relative to gravity are summarized in tables 1 and 2 respectively.
From
table 2 we see that the moduli forces can easily reach gravitational
strength. The strange-modulus $\phi_s$ for example mediates a force
of
gravitational strength at a distance of 0.5 mm. The magnitude of the
forces and Compton wavelengths are quite sensitive to the various
parameters
as well as the uncertainties in the matrix elements of the quark
currents.
For example, if $\lambda^{(1)}/\lambda^{(0)}=10$, the strength of
the moduli forces become 100 times larger and their range 10 times
smaller.
In particular the strange-modulus force is 100 times stronger than
gravity
and all other moduli-forces are also stronger than gravity.
Similarly,
if the scale $M$ is lowered to coincide with the grand unification
mass,
$M_{\rm GUT}=10^{16}$ GeV, then all moduli-forces increase by a
factor of
600. Typical ranges of $\lambda_\phi$ and ${\cal G}_\phi / 4 \pi$ for
several different $\phi$ are shown in fig.~2. The areas are obtained
by
taking $kN=1$, $\tan\beta=1$, $\lambda^{(2)} =0$, by varying
$\sqrt{F}$
between 30 and 100 TeV, and by varying $\lambda^{(0)}
/\lambda^{(1)}\times M/(5\times 10^{17}{\rm GeV})$ between $10^{-2}$
and $10^2$.

The dilaton couples to nucleons with a stength which can be as large as
80 times
gravity \cite{ven}. However since it couples to all fields in
the theory,
it is expected to receive a mass $\sim F/M$ from its strong
coupling
to the primordial supersymmetry-breaking sector. This would make its
Compton
wavelength less than $10^{-2}~\mu{\rm m}~(100~{\rm TeV})^2/F$, which
is too
short to be experimentally observed.
 
However another interesting possibility is the coupling of a modulus
to gluons. We are envisaging here a field $\phi$ which does not
universally
couple to all gauge bosons in the theory (in particular to a possible
strongly
interacting sector responsible for supersymmetry breaking), but only
to
the standard model gauge bosons. We assume an effective coupling
given by
\beq
{\cal L}=\frac{\lambda_g}{8\pi^2}\frac{\phi}{M}G_{\mu \nu}^aG^{a \mu
\nu}~.
\label{glue}
\eeq
Here $\lambda_g$ is an undetermined coupling constant and the factor
of
$8\pi^2$  is there to account for the fact that gauge couplings
depend
on moduli only at higher order.
 
Proceeding as before, we can estimate an upper bound on the $\phi$
Compton
wavelength by computing the contribution to its mass coming from the
interaction in eq.~(\ref{glue}),
\beq
m_\phi^2 =
k\frac{\alpha_s}{4\pi}\frac{\lambda_g^2}{(8\pi^2)^2}\frac{M_{\tilde
g}^2\Lambda^2}{M^2}~.
\eeq
The gluino mass $M_{\tilde g}$ is related to the
supersymmetry-breaking
scale $F$ by \cite{gaug,alv}
\beq
M_{\tilde g}=\frac{\alpha_s}{4\pi}N\frac{F}{\Lambda}~.
\eeq
The final expression for the Compton wavelength of the field $\phi$
is
\beq
\lambda_\phi = 8\times 10^{-4}~{\rm m} ~\left(\frac{M}{5\times
10^{17}~
{\rm GeV}}\right)~\frac{(100~{\rm TeV})^2}{F}~.
\eeq
 
With the help of eq.~(\ref{matq}), we obtain the following nucleon
matrix
element of the gluon operator:
\beq
\langle N | -\frac{9\alpha_s}{8\pi} G_{\mu \nu}^aG^{a \mu \nu}
|N\rangle = 0.8~m_N ~.
\eeq
Therefore the $\phi$-nucleon coupling can be still expressed by
eq.~(\ref{couppp}), where ${\cal G}_\phi$ is now given by
\beq
{\cal G}_\phi =\frac{\lambda_g}{8\pi^2}\frac{M_{\rm PL}}{M}
\frac{\langle N | G_{\mu \nu}^aG^{a \mu \nu}
|N\rangle }{m_N} \simeq -6~\lambda_g ~\left(\frac{5\times
10^{17}~{\rm GeV}}{M}
\right)~.
\eeq
Figure 2 shows the range of $\lambda_{\phi_g}$ and
${\cal G}_{\phi_g} / 4 \pi$ obtained by
taking $\sqrt{k}N=1$ and by varying $\sqrt{F}$
between 30 and 100 TeV and $\lambda_g^{-1}\times M/(5\times
10^{17}~{\rm GeV})$
between $10^{-2}$ and $10^2$.
 
\section{Pseudo-Goldstone Bosons}
 
In this section we consider long-range forces mediated by {\pgb}.
{\ppgb} originating from the spontaneous breaking of an {\it abelian}
symmetry
have only pseudoscalar couplings to fermionic matter fields. Thus
they
mediate spin-dependent forces with potential falling off with the
distance
$r$ as $1/r^3$, because, in the non-relativistic limit,
pseudoscalar couplings involve spin-flip transitions.
However long-range forces induced by scalar
couplings of the {\pgb} to the matter fields can be present if CP is
violated.
Examples of such an effect in the case of CP violation from the
Kobayashi-Maskawa matrix \cite{moody}, the $\theta$-term in QCD
\cite{chang},
or new kind of interactions \cite{ross} have already been presented
in the
literature.
 
Here we want to show that {\pgb} of {\it non-abelian} broken
symmetries can
mediate long-range forces, even in the absence of any CP violation.
Let us consider for instance the case of a unitary non-abelian
symmetry
and let us choose the generators to be either purely real or purely
imaginary.
From the interaction Lagrangian one observes that Goldstone bosons
corresponding to transformations along real (imaginary) generators
are
CP-odd (CP-even) particles. The CP symmetry does not exclude
therefore scalar
couplings of some of the {\pgb} to matter. These {\pgb}, even under
CP,
correspond to imaginary generators
and therefore only couple to matter fields with different unitary
symmetry
indices. Since the unitary symmetry is explicitly broken,
non-diagonal
couplings can be converted into diagonal couplings and ultimately
into
long-range coherent interactions.
 
We will
illustrate this phenomenon in a class of supersymmetric models
introduced
in ref.~\cite{dgt}, where the dynamics can solve the flavour
problem\footnote{The dynamical determination of the flavour
parameters is much
in the same spirit as the dynamical determination of the top-quark
mass,
proposed in ref.~\cite{kou}.},
{\it i.e.} it can appropriately suppress all flavour-changing neutral
currents. These theories have a global
$U(3)^5$ flavour symmetry group, which is spontaneously (but not
explicitly)
broken in the limit of vanishing Yukawa couplings. The corresponding
Goldstone bosons acquire small masses from the explicit source of
symmetry
breaking (the Yukawa couplings) and can potentially mediate
long-range
forces. The fundamental assumption underlying these theories
is that the soft supersymmetry-breaking masses for the different
squarks and
sleptons $\tilde{m}^2_A$ ($A=Q,\bar U ,\bar D ,L, \bar E$)
are promoted to fields:
\beq
\tilde{m}^2_A \to \Sigma_A \equiv U^\dagger_A \bar \Sigma_A U_A
~,
\label{sig}
\eeq
where $\bar \Sigma_A$ are diagonal matrices with real, positive
parameters ordered according to increasing magnitude.
Each $\bar \Sigma_A$ corresponds to the vacuum expectation value of
the field
$\Sigma_A$, which spontaneously break a $U(3)$ factor of the flavour
group.
$U_A$ are
$3\times 3$ unitary matrices in flavour space containing the Golstone
bosons
$\sigma^\alpha_A$. They can be written explicitly as
\beq
U_A
= {\rm exp}(i\sum_\alpha \lambda^\alpha\sigma^\alpha_A )~,
\label{up}
\eeq
where
$\lambda^\alpha $ are the Gell-Mann matrices. The sum in
eq.~(\ref{up})
extends over the generators of the flavour
group
broken by $\bar \Sigma_A$, in short the six generators of
$SU(3)/U(1)^2$.
The supersymmetry-breaking trilinear terms $A$ can also
be interpreted as fields
in an analogous fashion \cite{dgt}. Here we ignore these
terms for simplicity, as they
do not affect the essential properties of the Goldstone-mediated
long-range
forces.
 
In order to estimate the masses of the {\pgb}, we need to specify the
flavour
symmetry breaking part of the potential. This was done in
ref.~\cite{dgt},
under the assumption that the Yukawa couplings $h_e$, $h_u$, and
$h_d$
are the only source of explicit flavour breaking. The dominant
contribution
to the $\sigma_A^\alpha$ masses come from the following terms in the
effective potential:
\beq
V_{\rm eff}=-\frac{2}{(4\pi )^4}\Lambda^2 {\rm Tr}\left[
h_e^\dagger h_e \Sigma_L+h_e h_e^\dagger  \Sigma_{\bar E}+
\frac{7}{4}(h_u^\dagger h_u +h_d^\dagger  h_d)\Sigma_Q+
\frac{7}{4}h_u h_u^\dagger  \Sigma_{\bar U}+
\frac{7}{4}h_d h_d^\dagger  \Sigma_{\bar D}\right]~.
\label{veff}
\eeq
$\Lambda$ is the cut-off scale, which roughly
corresponds to the mass of the
messenger particles which communicate supersymmetry breaking to the
observable sector. Equation (\ref{veff}) exhibits a strong
sensitivity on the physics
near $\Lambda$ caused by a two-loop quadratic divergence in the
$\sigma^\alpha_A$
mass (or, in other words, in the zero-point energy of a conventional
softly-broken supersymmetric theory \cite{rand,dgt}).
 
Each term in eq.~(\ref{veff}) can be expanded in series of
$\sigma_A^\alpha$;
the first term gives, aside from a $\sigma_A^\alpha$-independent
constant,
\beq
{\rm Tr}(h_e^\dagger h_e \Sigma_L )=-\sum_\alpha (\sigma_L^\alpha )^2
\sum_{i>j} |\lambda^\alpha_{ij}|^2({\bar \Sigma}_{Li}-{\bar
\Sigma}_{Lj})
(h_{ei}^2-h_{ej}^2) + {\cal O}(\sigma_L^{\alpha 3})~,
\label{beta}
\eeq
and analogous expressions hold for the other terms.
Here $h_{ei}$ is the charged-lepton Yukawa coupling of the $i$-th
generation.
Equations (\ref{veff}) and (\ref{beta}) show explicitly
the dynamical alignment of the soft masses and Yukawa couplings in
flavour
space, as found in ref.~\cite{dgt}. Notice that as an effect of the
non-trivial Kobayashi-Maskawa matrix, $h_d$ and $h_u$ cannot be
simultaneously
diagonalized and the alignment in the left quark-squark sector
is not complete. Nevertheless
flavour-changing neutral current processes are adequately suppressed.
This provides a solution of the flavour problem in supersymmetric
theories
very different than the usual assumption of soft-terms universality.
Squark masses have here a high degree of non-degeneracy, but their
mixing
angles are closely related to the quark mixing angles\footnote{A
similar
solution of the flavour problem, based however on family symmetries
rather
than a dynamical principle, has been proposed in ref.~\cite{nir}.}.
 
Before we can identify the physical masses of the {\pgb},
we need to define the canonical fields.  Dimensional
analysis and
the hypothesis that flavour symmetry is broken spontaneously at a
scale
$f$
suggest that
the field $\Sigma'$ defined by
\beq
\Sigma = \frac{m^2_s}{f} \Sigma'
\label{sigp}
\eeq
is
canonically normalized. Here $f$ can be identified with the Planck
mass
or possibly with some lighter scale connected with flavour breakdown,
as
$M_{\rm GUT}$.
In eq.~(\ref{sigp})
$m_s$ is the typical mass scale of the soft supersymmetry-breaking
terms.
We want to stress however that our choice of $\Sigma'$ being the
canonical field is arbitrary and different choices can lead to
different
masses and couplings for the physical particles.
The properly normalized kinetic term is
\beq
\frac{1}{2}~{\rm Tr}~\partial_\mu \Sigma_A' \partial^\mu \Sigma_A' =
\left( \frac{f}{m_s^2}\right)^2
\sum_\alpha
\partial_\mu
\sigma^\alpha_A \partial^\mu \sigma^\alpha_A \sum_{i>j} \vert
\lambda^\alpha_{ij} \vert^2
(\bar \Sigma_{A_i}-
\bar \Sigma_{A_j})^2~.
\eeq
From this and eqs.~(\ref{veff}-\ref{beta}) we can read the physical
masses squared of
the $\sigma^\alpha_A$ particles\footnote{The non-linear
Goldstone parametrization used
here makes sense only if the explicit symmetry breaking
$(h_i^2-h_j^2)\langle
H\rangle^2$
is not larger than the spontaneous breaking ($\bar \Sigma_i-\bar
\Sigma_j$). This is why eq.~(\ref{ult}) apparently blows up as $\bar
\Sigma_i -\bar \Sigma_j \to 0$.}:
\beq
m^2_{\sigma^\alpha_A} = \frac{2c\Lambda^2 }{(4\pi)^4}~
\frac{ m^4_s}
{f^2}~\sum_{i>j}\vert
\lambda^\alpha_{ij} \vert^2 \left( \frac{h^2_i -
h^2_j}{\bar \Sigma_{A_i} - \bar \Sigma_{A_j}}\right)~,
\label{ult}
\eeq
where $h$ are the corresponding Yukawa couplings and $c=1$ for
leptons,
$c=7/4$ for quarks.
From eq.~(\ref{ult}), we see
that $\sigma_A^4, \sigma_A^5, \sigma_A^6, \sigma_A^7$ get masses
proportional to the
third generation fermion mass of species $A$, while  $\sigma_A^1,
\sigma_A^2$
get masses
proportional only to the second generation fermion mass of species
$A$.  A
convenient
expression for the masses of the $\sigma_A^\alpha$ ($A=Q,{\bar
U},{\bar D}$)
is
\beq
m_{\sigma^\alpha_A} \simeq 3\times 10^{-4}~{\rm eV}~\left(
\frac{10^{16}~{\rm GeV}}{f} \right)
\left( \frac{\Lambda}{100~{\rm TeV}}\right)
\left( \frac{m^A_{f\alpha}}{1~{\rm GeV}}
\right) \left(\frac{1/\sqrt{2}}{\cos\beta}\right)
 \left(
\frac{m_s}{300~{\rm GeV}} \right)
\sqrt{\frac{m^2_s}{\Delta m^2_s}}~.
\eeq
Here
$\Delta m^2_s/m_s^2$ is the relevant sparticle mass splittings and
  $m^A_{f_\alpha}$ denotes the
third (second) generation
fermion mass of species $A$ if $\alpha = 4, 5, 6, 7$ ($\alpha = 1,
2$).
The Compton wavelengths
of the $\sigma^\alpha$
particles are
\beq
\lambda_{\sigma^\alpha_A} \simeq 6\times 10^{-4}~{\rm m}~\left(
\frac{f}{10^{16}~{\rm GeV}} \right)
\left( \frac{100~{\rm TeV}}{\Lambda}\right)
\left( \frac{1~{\rm GeV}}{m^A_{f\alpha}}
\right) \left(\frac{\cos\beta}{1/\sqrt{2}}\right)
 \left(
\frac{300~{\rm GeV}}{m_s} \right)
\sqrt{\frac{\Delta m^2_s}{m^2_s}}~.
\label{compt}
\eeq
Thus, unless the squark mass splitting is very small, the $\sigma$
can mediate forces between two
objects
separated by a macroscopic distance and lead to deviations from the
equivalence
principle.
 
Let us now turn to discuss the couplings of the {\pgb} to matter. As
we have previously discussed,
the CP properties of the $\sigma^\alpha$
are essential in determining the nature of the
forces. The $\sigma^\alpha$ that correspond to imaginary
$\lambda^\alpha$, namely $\sigma_2, \sigma_5$ and $\sigma_7$, are
CP-even
scalars and
they can mediate $1/r^2$ forces.
Diagonal couplings of $\sigma_2, \sigma_5$ and $\sigma_7$
to ordinary matter will arise because of the mismatch between mass
and interaction eigenstates.
Since for leptons the mixing angles vanish, there are no diagonal
long-range forces coupled to lepton number.
Within our approximation of neglecting left-right squark mixings,
these
forces can only be mediated by $\sigma_Q$, as flavour violation
resides
in the left quark-squark sector.
 
We will work in the basis defined by $h_u={\hat h}_u$ and
$h_d={\hat h}_dK^\dagger$, where ${\hat h}_{u,d}$ are diagonal real
matrices
and $K$ is the unitary Kobayashi-Maskawa matrix. This basis is
particularly convenient because it approximately corresponds to the
mass
eigenbasis for all squarks\footnote{This is true unless the
splittings
of the squarks soft masses are smaller than the corresponding quark
mass
splittings. Here we are interested in a case where there is a
significant departure from universality.}.
 The coupling of the properly normalized $\sigma$ to squarks $\phi$
is given
by
\beq
{\cal L}_{\sigma  \phi \phi} = \frac{i}{\sqrt 2} \frac{m^2_s}{f}
\sum_{\alpha}\sum_{i>j}
\phi^\star_i \lambda^\alpha_{i,j}\phi_j\sigma^\alpha + {\rm h.c.}
\label{coup}
\eeq

The interactions of the {\pgb}
with squarks,  eq.~(\ref{coup}), can be converted into a
diagonal coupling to ordinary matter exploiting the Kobayashi-Maskawa
angles which rotate the down quarks from the basis we are working
into
their mass eigenbasis. This can be done via one-loop diagrams
mediated
either by gluinos (for the coupling to down quarks) or by charginos
(for both up and down quarks). It is reasonable to expect that strong
interactions make the gluino exchange dominant over the chargino,
although
this may depend on the various parameters. The gluino-exchange
produces
a scalar effective coupling between $\sigma_Q^\alpha$ and a pair of
down quarks $d_k$, with identical flavour index $k$, given by
\beq
{\cal L}_{\sigma \bar d d} =  \frac{\sqrt{2}\alpha_s}{9 \pi}
\frac{m^2_s}{M_{\tilde g}^2} \frac{m_{d_k}}{f} ~\bar d_k d_k
\sigma_Q^\alpha
\sum_{i>j}{\rm Im}(\lambda^\alpha_{ij}K_{jk} K^\star_{ik})~
g\left( \frac{\bar \Sigma_i}{M_{\tilde g}^2} ,
\frac{\bar \Sigma_j}{M_{\tilde g}^2} \right)
~,
\label{neweq}
\eeq
where
\beq
g(x,y)=\frac{3}{2(x-y)}\left[
\frac{1}{x-1}+\frac{x(x-2)}{(x-1)^2}\log x
-(x \to y)\right]
\eeq
is normalized so that $g(1,1)=1$ and $M_{\tilde g}$ is the gluino
mass.
It is apparent from eq.~(\ref{neweq})
that if CP is conserved, or in other words if $K$ is real,
only imaginary $\lambda^\alpha$ can generate scalar couplings.
Equation
(\ref{neweq}) is also proportional to the down quark mass, $m_{d_k}$.
This was to be expected because, as
$m_{d_k} \to 0$, $K$ looses its meaning and no flavour transitions
are allowed.
 
The effective $\sigma_Q$-nucleon
coupling is
\beq
{\cal L}_{\sigma_Q^\alpha \bar N N} =
\frac{m_N }{M_{\rm PL}} {\cal G}_{\sigma_Q^\alpha} \sigma_Q^\alpha
\bar \psi_N \psi_N~,
  \eeq
where
${\cal G}_{\sigma_Q^\alpha}$ measures the strength
of the $\sigma_Q^\alpha$-coupling relative to gravity\footnote{We
have
neglected here an effective $\sigma_Q^\alpha$-gluon coupling coming
from the
integration of a squark loop. Its contribution is expected to be
smaller
than the contribution given in eq.~(\ref{neweq}).},
\beq
{\cal G}_{\sigma^\alpha} = \frac{ \sqrt{2}\alpha_s}{9 \pi}
\frac{m^2_s}{M_{\tilde g}^2} \frac{M_{\rm PL}}{f}\sum_k
\sum_{i>j}{\rm Im}(\lambda^\alpha_{ij}K_{jk} K^\star_{ik} )~
g\left( \frac{\bar \Sigma_i}{M_{\tilde g}^2} ,
\frac{\bar \Sigma_j}{M_{\tilde g}^2} \right)~\frac{\langle N |
m_{d_k}{\bar d}_k d_k |N\rangle}{m_N}
~.
\eeq
 
The Goldstone forces' range and strength relative to gravity are
shown in
tables 1 and 2 respectively. The Goldstone forces are significantly
weaker than the moduli forces because of the associated global
symmetries. As a result prehaps the only Goldstone that has a chance
to
be observable is $\sigma_Q^2$, provided that the flavour scale $f$ is
around $M_{\rm GUT}$ or lower.
 
Finally we want to recall that, in simple cosmologies, the
Goldstones as well as the moduli suffer from the usual "cosmological
moduli problem" \cite{moduli}. We have nothing to add to that
except to hope that either inflation or some other mechanism
solves this problem.
 
\section{Prospects}
 
Present limits on new forces at scales larger than 1 cm are reviewed
in ref.~\cite{lrf} .
Existing experimental limits at shorter distances come from two
sources, the
electromagnetic Casimir force measurements \cite{casimir} and the 2
cm
Cavendish experiment \cite{cavendish} which are shown in fig.~2.
From this we see that the Casimir force measurement
allows for new forces up to $10^9$ times gravity in the range
between $\sim 10^{-4}$ cm and
$\sim 10^{-1}$ cm.
This constraint is a steep function of distance and becomes
even weaker at short distances where the Casimir force is larger; at
10 $\mu$m it allows a force up to $10^{12}$ times gravity.
The Cavendish experiment at 2 cm gives a constraint that decays
exponentially in significance at distances below a cm; at 0.5 mm it
allows a new force up to $10^{8}$ times gravity.
Cryogenic mechanical oscillator techniques have been proposed
\cite{cryogenic}
that can improve the existing limits by up to $10^{10}$ in the range
between 40 microns and 1 cm  and can detect forces $10^{-2}$ times
gravity
with a range greater than 40 microns. The dotted lines in fig.~2
indicate the sensitivity of these techniques. The region to the left
of the steep dotted line is inaccessible because the background
electrostatic force from the surface potential dominates. The region
below the dotted line is swamped by the Newtonian background due to
edge effects arising from the finite size of the parallel plates.
Another  proposal under consideration involves atomic beams
\cite{Kasevich}; it is not shown in the figure because its domain of
sensitivity is still being studied. In the figure we also show bands
that correspond to some of our predictions as we vary the unknown
parameters of our theory. We see that a broad range is accessible to
the cryogenic oscillator techniques.
 
To summarize, there are two essential ingredients that led to the
qualitative conclusions of this paper. First, the existence of
scalars with gravitational couplings which get mass only from
supersymmetry breaking. Second, the hypothesis that supersymmetry
breaking originates at low energies, not too far from the weak scale.
These simple hypotheses led us to the possibilities discussed here.
Of course, the numerical uncertainties associated with the magnitude
and range of the forces are large. Nevertheless, we hope that these
estimates will motivate renewed efforts for searches of new sub-cm
forces. It seems hard to overestimate the importance of discovering
such a force. It would provide us with a rare
window into Planckean Physics  and the scale of supersymmetry
breaking. It is even possible that a study of the
material and distance dependence of these forces could give us a more
detailed picture of how flavour symmetry emerges from the Planck
scale.

\section{Acknowledgements}
It is a pleasure to thank Costas Kounnas, Michael Peskin,
Leonard Susskind, and Gabriele Veneziano for
valuable discussions and our experimental colleagues  Steve Chu, Mark
Kasevich, Peter Michelson and Doug Osheroff for teaching us about the
experimental feasibility of testing these ideas over the last few
months.
 
 \vfill\eject

\def\ijmp#1#2#3{{\it Int. Jour. Mod. Phys. }{\bf #1~}(19#2)~#3}
\def\pl#1#2#3{{\it Phys. Lett. }{\bf B#1~}(19#2)~#3}
\def\zp#1#2#3{{\it Z. Phys. }{\bf C#1~}(19#2)~#3}
\def\prl#1#2#3{{\it Phys. Rev. Lett. }{\bf #1~}(19#2)~#3}
\def\rmp#1#2#3{{\it Rev. Mod. Phys. }{\bf #1~}(19#2)~#3}
\def\prep#1#2#3{{\it Phys. Rep. }{\bf #1~}(19#2)~#3}
\def\pr#1#2#3{{\it Phys. Rev. }{\bf D#1~}(19#2)~#3}
\def\np#1#2#3{{\it Nucl. Phys. }{\bf B#1~}(19#2)~#3}
\def\mpl#1#2#3{{\it Mod. Phys. Lett. }{\bf #1~}(19#2)~#3}
\def\arnps#1#2#3{{\it Annu. Rev. Nucl. Part. Sci. }{\bf
#1~}(19#2)~#3}
\def\sjnp#1#2#3{{\it Sov. J. Nucl. Phys. }{\bf #1~}(19#2)~#3}
\def\jetp#1#2#3{{\it JETP Lett. }{\bf #1~}(19#2)~#3}
\def\app#1#2#3{{\it Acta Phys. Polon. }{\bf #1~}(19#2)~#3}
\def\rnc#1#2#3{{\it Riv. Nuovo Cim. }{\bf #1~}(19#2)~#3}
\def\ap#1#2#3{{\it Ann. Phys. }{\bf #1~}(19#2)~#3}
\def\ptp#1#2#3{{\it Prog. Theor. Phys. }{\bf #1~}(19#2)~#3}

\vfill\eject

\begin{table}
\caption[]{Estimates for the Compton wavelengths of the light scalar
particles
discussed in the text.}
\vglue.3cm
\begin{center}
\begin{tabular}{|c|cc|}
\hline
Particle & $\lambda$~~[m] & \\
\hline
\hline
 & & \\
$\phi_u$ & $2\times 10^{-2}$ &
$\frac{(100~{\rm TeV})^2}{F}~
\left( \frac{M}{5\times 10^{17}~{\rm GeV}}\right)~
\left( \frac{\sin\beta}{1/\sqrt{2}}\right)
\left[ kN
\left(\frac{\lambda_u^{(1)2}}{\lambda_u^{(0)2}}
+\frac{\lambda_u^{(2)}}{\lambda_u^{(0)}}\right) \right]^{-1/2}$\\
$\phi_d$ & $1\times 10^{-2}$ &
$\frac{(100~{\rm TeV})^2}{F}~
\left( \frac{M}{5\times 10^{17}~{\rm GeV}}\right)~
\left( \frac{\cos\beta}{1/\sqrt{2}}\right)
\left[ kN
\left(\frac{\lambda_d^{(1)2}}{\lambda_d^{(0)2}}
+\frac{\lambda_d^{(2)}}{\lambda_d^{(0)}}\right) \right]^{-1/2}$\\
$\phi_s$ & $5\times 10^{-4}$ &
$\frac{(100~{\rm TeV})^2}{F}~
\left( \frac{M}{5\times 10^{17}~{\rm GeV}}\right)~
\left( \frac{\cos\beta}{1/\sqrt{2}}\right)
\left[ kN
\left(\frac{\lambda_s^{(1)2}}{\lambda_s^{(0)2}}
+\frac{\lambda_s^{(2)}}{\lambda_s^{(0)}}\right) \right]^{-1/2}$\\
$\phi_c$ & $7\times 10^{-5}$ &
$\frac{(100~{\rm TeV})^2}{F}~
\left( \frac{M}{5\times 10^{17}~{\rm GeV}}\right)~
\left( \frac{\sin\beta}{1/\sqrt{2}}\right)
\left[ kN
\left(\frac{\lambda_c^{(1)2}}{\lambda_c^{(0)2}}
+\frac{\lambda_c^{(2)}}{\lambda_c^{(0)}}\right) \right]^{-1/2}$\\
$\phi_b$ & $2\times 10^{-5}$ &
$\frac{(100~{\rm TeV})^2}{F}~
\left( \frac{M}{5\times 10^{17}~{\rm GeV}}\right)~
\left( \frac{\cos\beta}{1/\sqrt{2}}\right)
\left[ kN
\left(\frac{\lambda_b^{(1)2}}{\lambda_b^{(0)2}}
+\frac{\lambda_b^{(2)}}{\lambda_b^{(0)}}\right) \right]^{-1/2}$\\
$\phi_t$ & $5\times 10^{-7}$ &
$\frac{(100~{\rm TeV})^2}{F}~
\left( \frac{M}{5\times 10^{17}~{\rm GeV}}\right)~
\left( \frac{\sin\beta}{1/\sqrt{2}}\right)
\left[ kN
\left(\frac{\lambda_t^{(1)2}}{\lambda_t^{(0)2}}
+\frac{\lambda_t^{(2)}}{\lambda_t^{(0)}}\right) \right]^{-1/2}$\\
 & & \\
\hline
 & & \\
$\phi_g$ & $8\times 10^{-4}$ &
$\frac{(100~{\rm TeV})^2}{F}~
\left( \frac{M}{5\times 10^{17}~{\rm GeV}}\right)~
\left( \sqrt{k}N \lambda_g \right)^{-1}$\\
 & & \\
\hline
 & & \\
$\sigma_Q^2$ & $4\times 10^{-3}$ &
$\left(
\frac{f}{10^{16}~{\rm GeV}} \right)
\left( \frac{100~{\rm TeV}}{\Lambda}\right)
\left(\frac{\cos\beta}{1/\sqrt{2}}\right)
 \left(
\frac{300~{\rm GeV}}{m_s} \right)
\sqrt{\frac{\Delta m^2_s}{m^2_s}}$\\
$\sigma_Q^5$ & $1\times 10^{-4}$ &
$\left(
\frac{f}{10^{16}~{\rm GeV}} \right)
\left( \frac{100~{\rm TeV}}{\Lambda}\right)
\left(\frac{\cos\beta}{1/\sqrt{2}}\right)
 \left(
\frac{300~{\rm GeV}}{m_s} \right)
\sqrt{\frac{\Delta m^2_s}{m^2_s}}$\\
$\sigma_Q^7$ & $1\times 10^{-4}$ &
$\left(
\frac{f}{10^{16}~{\rm GeV}} \right)
\left( \frac{100~{\rm TeV}}{\Lambda}\right)
\left(\frac{\cos\beta}{1/\sqrt{2}}\right)
 \left(
\frac{300~{\rm GeV}}{m_s} \right)
\sqrt{\frac{\Delta m^2_s}{m^2_s}}$\\
 & & \\
\hline
\end{tabular}
\end{center}
\end{table}

\vfill\eject

\begin{table}
\caption[]{Estimates for the couplings relative to gravity of the new
forces mediated by the light scalar particles
discussed in the text.}
\vglue0.3cm
\begin{center}
\begin{tabular}{|c|cc|}
\hline
Particle & ${\cal G}^2/(4\pi )$ &  \\
\hline
\hline
 & & \\
$\phi_u$ & $2\times 10^{-2}$ &
$\left(\frac{5\times 10^{17}~{\rm GeV}}{M}\right)^2
\left( \frac{\lambda_u^{(1)}}{\lambda_u^{(0)}}\right)^2$\\
$\phi_d$ & $8\times 10^{-2}$ &
$\left(\frac{5\times 10^{17}~{\rm GeV}}{M}\right)^2
\left( \frac{\lambda_d^{(1)}}{\lambda_d^{(0)}}\right)^2$\\
$\phi_s$ & $1$ &
$\left(\frac{5\times 10^{17}~{\rm GeV}}{M}\right)^2
\left( \frac{\lambda_s^{(1)}}{\lambda_s^{(0)}}\right)^2$\\
$\phi_c$ & $2\times 10^{-1}$ &
$\left(\frac{5\times 10^{17}~{\rm GeV}}{M}\right)^2
\left( \frac{\lambda_c^{(1)}}{\lambda_c^{(0)}}\right)^2$\\
$\phi_b$ & $2\times 10^{-1}$ &
$\left(\frac{5\times 10^{17}~{\rm GeV}}{M}\right)^2
\left( \frac{\lambda_b^{(1)}}{\lambda_b^{(0)}}\right)^2$\\
$\phi_t$ & $2\times 10^{-1}$ &
$\left(\frac{5\times 10^{17}~{\rm GeV}}{M}\right)^2
\left( \frac{\lambda_t^{(1)}}{\lambda_t^{(0)}}\right)^2$\\
 & & \\
\hline
 & & \\
$\phi_g$ & 3 &
$\left(\frac{5\times 10^{17}~{\rm GeV}}{M}\right)^2
\lambda_g^2$\\
 & & \\
\hline
 & & \\
$\sigma_Q^2$ & $5\times 10^{-3}$ &
$\left(\frac{10^{16}~{\rm GeV}}{f}\right)^2
\left[ \frac{m_s^2}{M_{\tilde g}^2}~g\left(
\frac{\bar \Sigma_2}{M_{\tilde g}^2},
\frac{\bar \Sigma_1}{M_{\tilde g}^2}\right) \right]^2$\\
$\sigma_Q^5$ & $5\times 10^{-6}$ &
$\left(\frac{10^{16}~{\rm GeV}}{f}\right)^2
\left[ \frac{m_s^2}{M_{\tilde g}^2}~g\left(
\frac{\bar \Sigma_3}{M_{\tilde g}^2},
\frac{\bar \Sigma_1}{M_{\tilde g}^2}\right) \right]^2$\\
$\sigma_Q^7$ & $5\times 10^{-5}$ &
$\left(\frac{10^{16}~{\rm GeV}}{f}\right)^2
\left[ \frac{m_s^2}{M_{\tilde g}^2}~g\left(
\frac{\bar \Sigma_3}{M_{\tilde g}^2},
\frac{\bar \Sigma_2}{M_{\tilde g}^2}\right) \right]^2$\\
 & & \\
\hline
\end{tabular}
\end{center}
\end{table}

\vfill\eject

\begin{figure}
\hglue3.5cm
\epsfig{figure=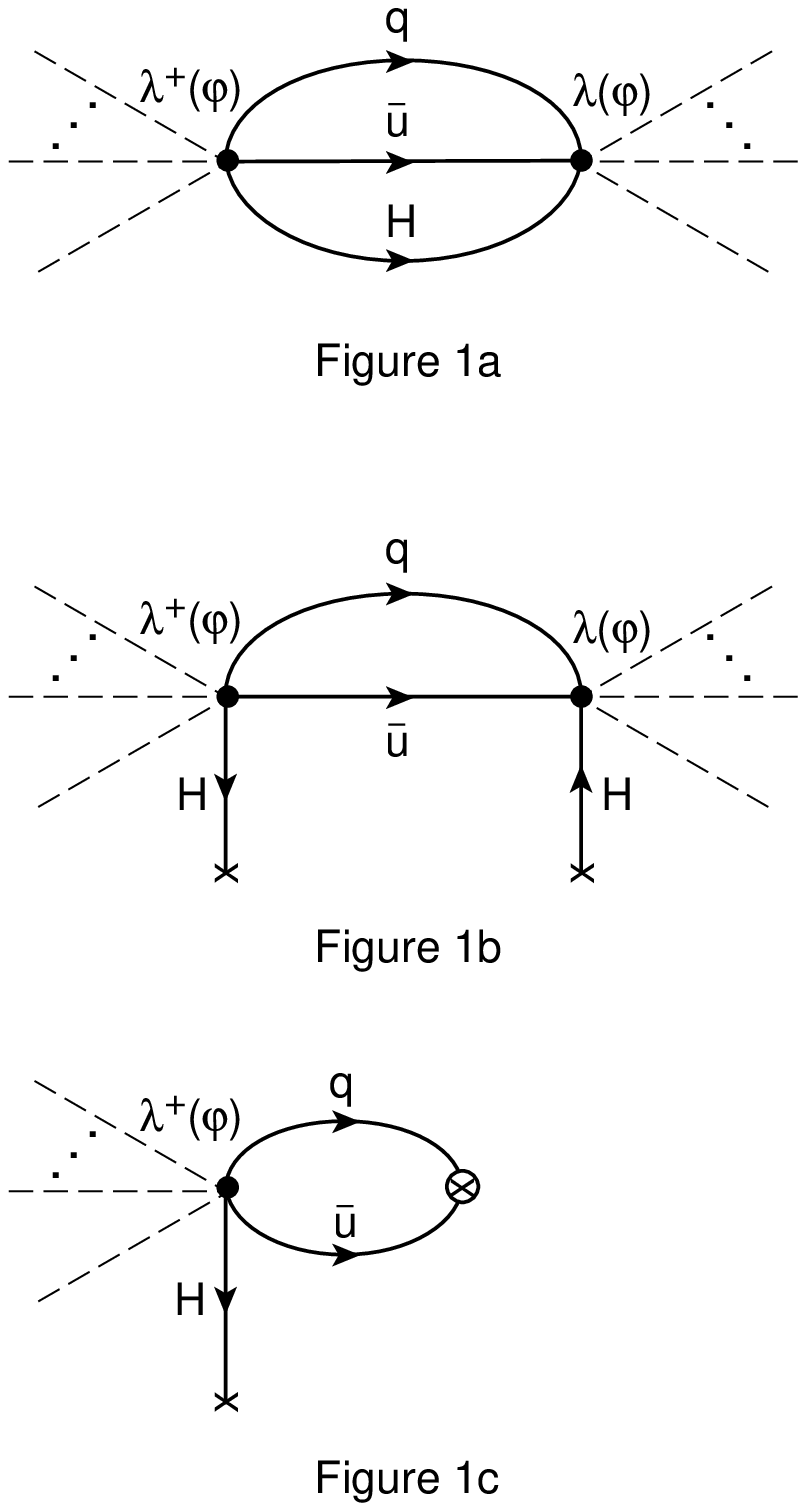,width=10cm}
\caption[]{Feynman diagrams contributing to the effective
potential
$V(\phi )$. The black dot represents the effective $\lambda$
interaction
and the cross represents the QCD condensate.}
\end{figure}
\begin{figure}
\hglue0.75cm
\epsfig{figure=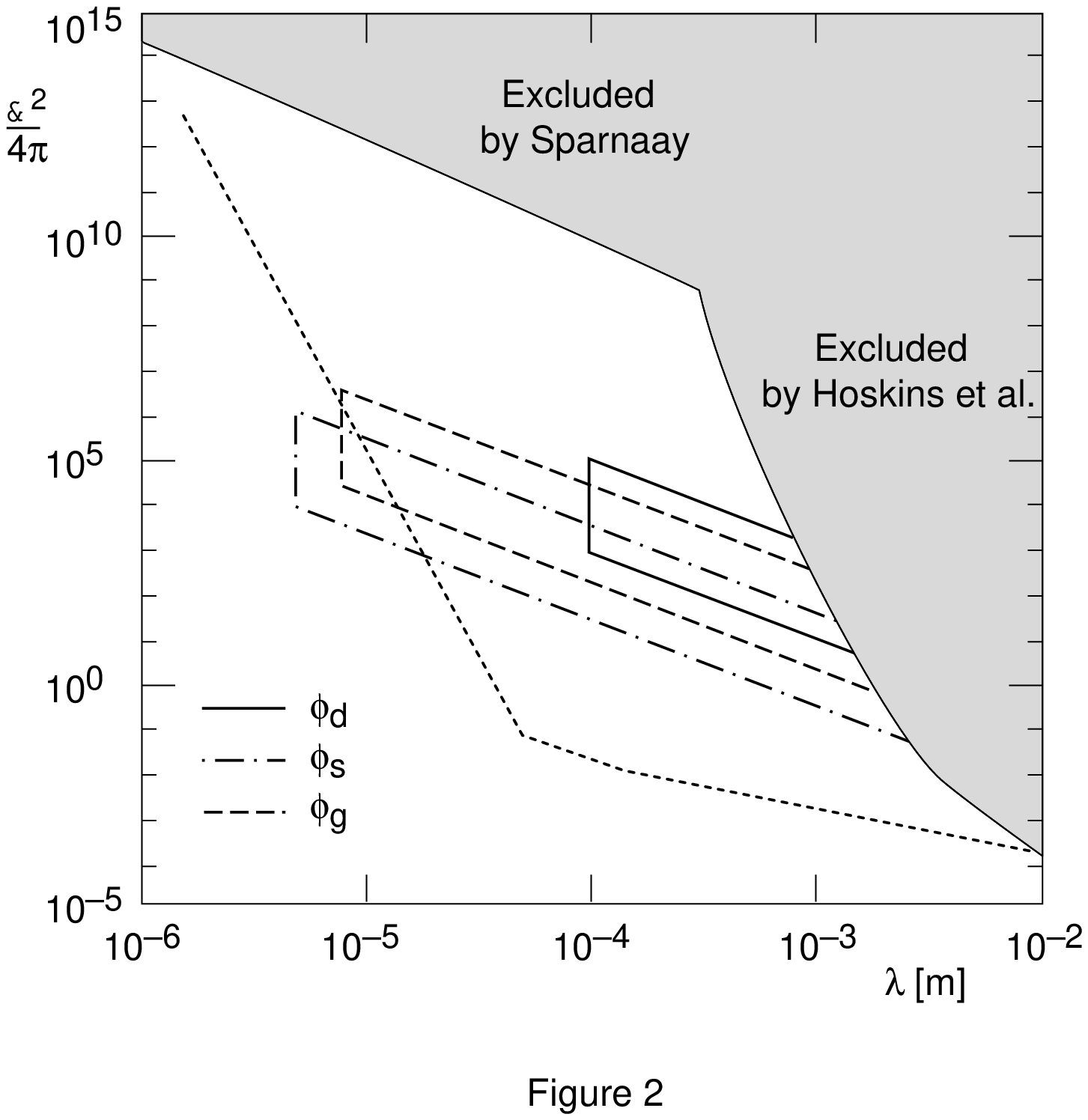,width=15.5cm}
\caption[]{Estimates for the region of Compton wavelength ($\lambda$)
and force strength relative to gravity (${\cal G}/4\pi$) corresponding
to strange- and down-Yukawa moduli ($\phi_s,\phi_d$) and gluon modulus
($\phi_g$).
The shaded region is experimentally  excluded by the searches
described
in refs.~\cite{casimir,cavendish}. The dotted line shows the expected
sensitivity of the planned experiment described in
ref.~\cite{cryogenic}.}
\end{figure}

\begin{thebibliography}{99}
 
\bibitem{ds} S.~Dimopoulos and D.~Sutter, \np{452}{95}{496}.
 
\bibitem{nil} H.P.~Nilles, \prep{110}{84}{1}.
 
\bibitem{gaug} M.~Dine, A.E.~Nelson, and Y.~Shirman,
\pr{51}{95}{1362};\\
M.~Dine, A.E.~Nelson, Y.~Nir, and Y.~Shirman, preprint SCIPP-95-32
(1995).
 
\bibitem{alv} L.~Alvarez-Gaum\'e, M.~Claudson, and M.~Wise,
\np{207}{82}{96}.
 
\bibitem{cina} T.P.~Cheng, \pr{38}{88}{2869};\\
H.-Y.~Cheng, \pl{219}{89}{347}.
 
 
\bibitem{leut} J.~Gasser, H.~Leutwyler, and M.E.~Sainio,
\pl{253}{91}{252}.
 
\bibitem{giap} T.~Hatsuda and T.~Kunihiro, \np{387}{92}{705}.
 
\bibitem{shif}
M.A. Shifman, A.I. Vainshtein, and V.I. Zakharov, {\it Phys. Lett.}
{\bf 78B} (1978) 443.
 
\bibitem{ven} G.~Veneziano and T.~Taylor, \pl{213}{88}{450}.
 
\bibitem{moody} J.E.~Moody and F.~Wilczek, \pr{30}{84}{130}.
 
\bibitem{chang} D.~Chang, R.~Mohapatra, and S.~Nussinov,
\prl{55}{85}{2835}.
 
\bibitem{ross} C.T.~Hill and G.G.~Ross, \pl{203}{88}{125} and
{\it Nucl. Phys.} {\bf 311} (1988/89) 253.
 
\bibitem{dgt} S.~Dimopoulos, G.F.~Giudice, and N.~Tetradis,
\np{454}{95}{59}.
 
\bibitem{kou} C. Kounnas, F. Zwirner, and I. Pavel, {\it Phys. Lett.}
{\bf B335}
(1994) 403;\\
P. Binetruy and E. Dudas, {\it Phys. Lett.} {\bf B338} (1994) 23.
 
\bibitem{rand} J.~Bagger, E.~Poppitz, and L. Randall,
\np{455}{95}{59}.
 
\bibitem{nir} Y.~Nir and N.~Seiberg, \pl{309}{93}{337}.
 
\bibitem{moduli} G. Coughlan, W. Fischler, E. Kolb, S. Raby, and G.
Ross,
{\it Phys. Lett.}
{\bf B131} (1983) 59;\\
 J .Ellis, D.V. Nanopoulos, and M. Quir\'os, {\it
Phys. Lett.}
{\bf B174} (1986) 176;\\
 G. German and G.G. Ross, {\it Phys. Lett.}, {\bf
B172}
(1986) 305;\\
 O. Bertolami, {\it Phys. Lett} {\bf B209} (1988) 277;\\
B. de Carlos, J.A. Casas, F. Quevedo, and E. Roulet, {\it Phys.
Lett.}
{\bf B318} (1993) 447;\\
T. Banks, D. Kaplan, and A. Nelson {\it Phys. Rev.}, {\bf D49} (1994)
779;\\
T. Banks, M. Berkooz, and P.J .Steinhardt, preprint RU-94-92.
 
\bibitem{lrf}
A.~De R\'ujula, \pl{180}{86}{213};\\
F.D.~Stacy {\it et al.}, \rmp{59}{87}{157};\\
J.~Ellis, N.C.~Tsamis, and M.~Voloshin, \pl{194}{87}{291};\\
C.~Talmadge {\it et al.}, \prl{61}{88}{1159};\\
V.L.~Fitch, M.V.~Isaila and M.A.~Palmer, {\it Phys. Rev.
Lett.} {\bf 60} (1988) 1801;\\
W.~Buchm\"uller, in {\it Proc. of the 8$^{th}$ Eloisatron Workshop on
Higgs Physics} (1989);\\
 B.~Heckel {\it et.~al.}, {\it Phys. Rev. Lett.} {\bf 63}  (1989)
2705.
 
\bibitem{casimir} M.J.~Sparnaay, {\it Physica} {\bf 24} (1958) 751.
 
\bibitem{cavendish} J.K.~Hoskins {\it et al.}, \pr{32}{85}{3084}.
 
\bibitem{cryogenic} J.C.~Price, in {\it Proc. Int. Symp. on
Experimental
Gravitational
Physics}, ed. by P.F.~Michelson, Guangzhou, China, World Scientific
(1988).
 
\bibitem{Kasevich} M.~Kasevich, private communication.
 
\end{thebibliography}
\end{document}